\documentclass[twoside]{report}
\usepackage{iwsm}
\usepackage{graphicx}
\usepackage{amsmath, amssymb}
\usepackage{booktabs}

\begin{document}


\title{Statistical models for dynamics in extreme value processes}
\titlerunning{Statistical models for dynamics in extreme value processes}

\author{Bernhard Spangl\inst{1},
        Sascha Desmettre\inst{2},
        Peter Ruckdeschel\inst{3}}
\authorrunning{Spangl et al.}

\institute{Institute\ of Applied Statistics and Computing, BOKU, Vienna, Austria
\and Department\ of Mathematics, TU Kaiserslautern, Germany
\and Institute for Mathematics, University Oldenburg, Germany
}

\email{bernhard.spangl@boku.ac.at}

\abstract{
We study four different
approaches to model time-dependent extremal behavior: dynamics introduced
by (a) a state-space model (SSM), (b) a shot-noise-type process with
GPD marginals, (c) a 
copula-based autoregressive model with GPD marginals,
and (d) a GLM with GPD marginals (and previous
extremal events as regressors).
Each of the models is fit against data, and from the fitted data,
we simulate corresponding paths according to the respective fitted models.
At this simulated data, the respective dependence structure is analyzed
in copula plots and judged against its capacity to fit the corresponding
inter-arrival distribution.}


\maketitle



\section{Motivation and issues} 


A challenge in dealing with extreme events in river discharge
data is to capture well time dynamics of these extremes, in particular
in the presence of seasonal effects and trends.



We will provide models which are able to capture the extreme
behaviour and provide simple and parsimonious, but yet flexible dynamics.

The goal is to be able to assess the magnitude of extreme events
as well as the inter-arrival time distribution of those extremes by
simulation (see also Khaliq\ et\ al.,\ 2006).


\section{Data generating processes}

To address these issues, we  discuss four different approaches to
incorporate dynamics into extreme value processes. For different choices
of parameters, each of these approaches is illustrated by typical
realizations to assess the induced dynamics and by lagged PP plots
to grasp the respective dependence structures.

\subsection{SSM approach}

As a first approach we propose to use the following model:
$$
X_t=\mu_t +\sigma_t v_t \ ,
$$
with
$$
\begin{array}{rcl}
\mu_t & = & \mu_0+\varphi X_{t-1} \ ,\\
\log \sigma_t & = &
                  \log \sigma_0 
                  + \gamma 
                  \log(1+(X_{t-1}-\mu_{t-1})^2) \ .
\end{array}
$$
This model is comparable to an AR-EARCH one and can be easily extended
to an ARMA-EGARCH model or other even more complex ones.

The iid innovations $v_t$ are generated according to the following
two-step procedure. First, we choose $v_t' \sim F_0$, where, e.g.,
$F_0 = {\mathcal N}(0, 1)$. Then, given a certain threshold $\tau$
the innovations are enriched by $v_t'' \sim {\rm GPD}$, i.e.,
\begin{displaymath}
v_t=\left\{\begin{array}{l@{\quad:\quad}l}
v'_t & |v'_t| < \tau \ ,\\
v_t''\sim{\rm GPD}(\mu=0,\sigma=1,\xi) & \mbox{else} \ .
\end{array}\right.
\end{displaymath}
$\xi$ is chosen according to the maximum domain of attraction of $F_0$.

\begin{figure}[bt!]\centering
\scalebox{0.35}{
\includegraphics{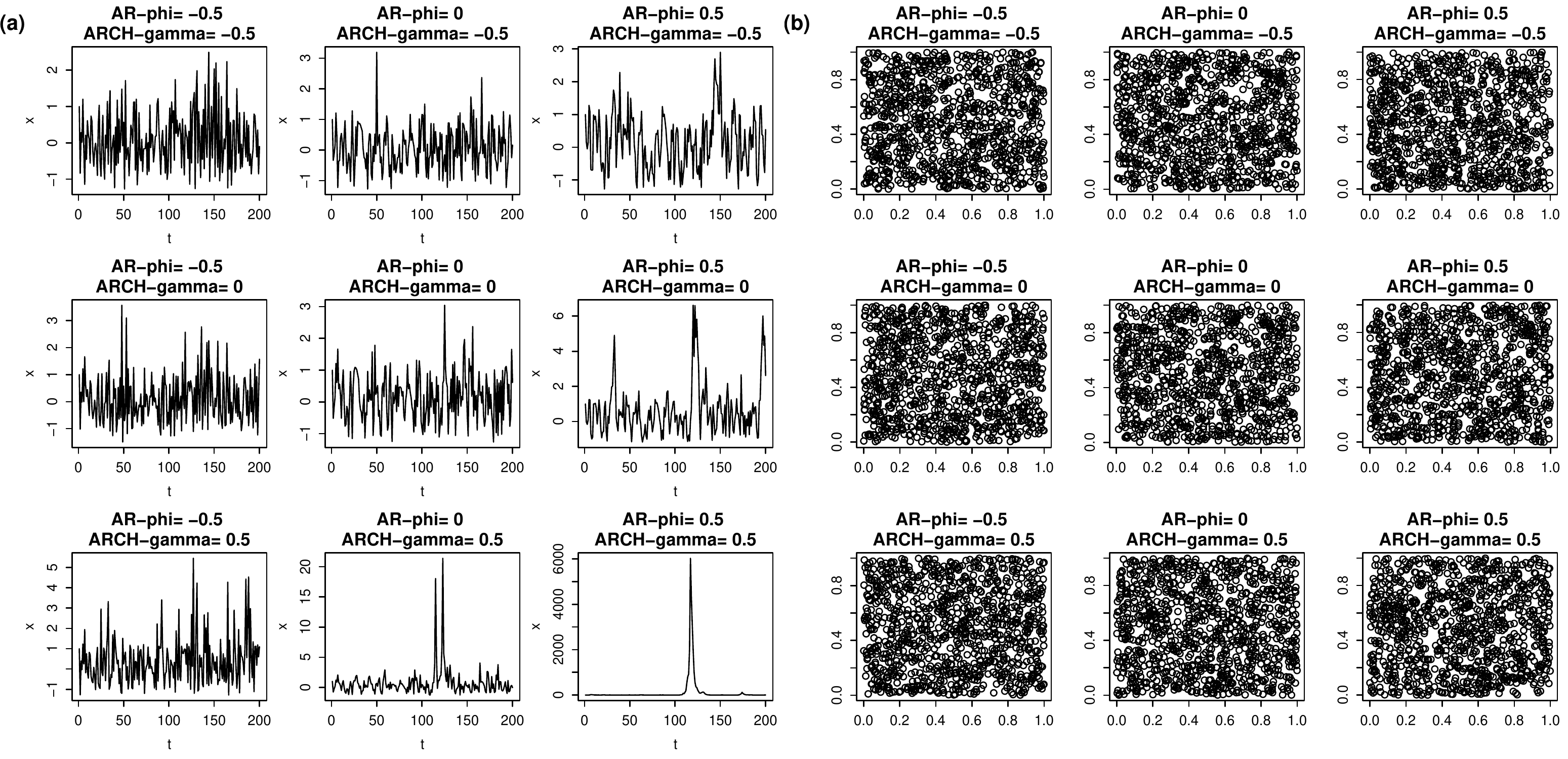}
}
\caption{\label{spangl:GPDSSM} SSM approach: (a) typical realizations,
and (b) lagged PP plots.}
\end{figure}

This state-space model approach allows for separate estimation of
time-dependency (via filtering) and of GPD-parameters, $\mu$, $\sigma$,
and $\xi$, of the marginal distribution of the innovations.
Typical realizations are shown in Figure\ \ref{spangl:GPDSSM}, together
with the corresponding lagged PP plots.

\subsection{Shot-noise-type approach}

As a second approach we suggest to use the following shot-noise-type
process:
$$
X_t = U_t Y_t + (1-U_t) \min\{v_t,Y_t\} \ ,
$$
where $U_t\sim {\rm Bin}(0,\beta)$ and $v_t\sim G={\rm GPD}(\mu,\sigma,\xi)$.
Let $G$ and $G^{-1}$ also denote the corresponding cdf and quantile
function, respectively.
Moreover, $\{U_t\}_t$ and $\{v_t\}_t$ are stochastically independent, and
$(U_t,v_t)$ is independent of $\{X_s\}_{s<t}$.
Then $Y_t=f_\beta(X_{t-1})$ with
$$
\begin{array}{rcl}
f_\beta & = & G^{-1}\circ f^0_\beta \circ G \ ,\\
f^0_\beta & = & {u}\big/{[(1-u)\beta+u]}\ .
\end{array}
$$
Hence, $X_t\sim G$. The parameter $\beta \in [0, 1]$ controls the dependency.
Again, this approach allows for separate estimation of time-dependency
(number of  dropdowns) and of GPD-parameters,
$\mu$, $\sigma$, $\xi$, of the marginal distribution.

\begin{figure}[bt!]\centering
\scalebox{0.35}{
\includegraphics{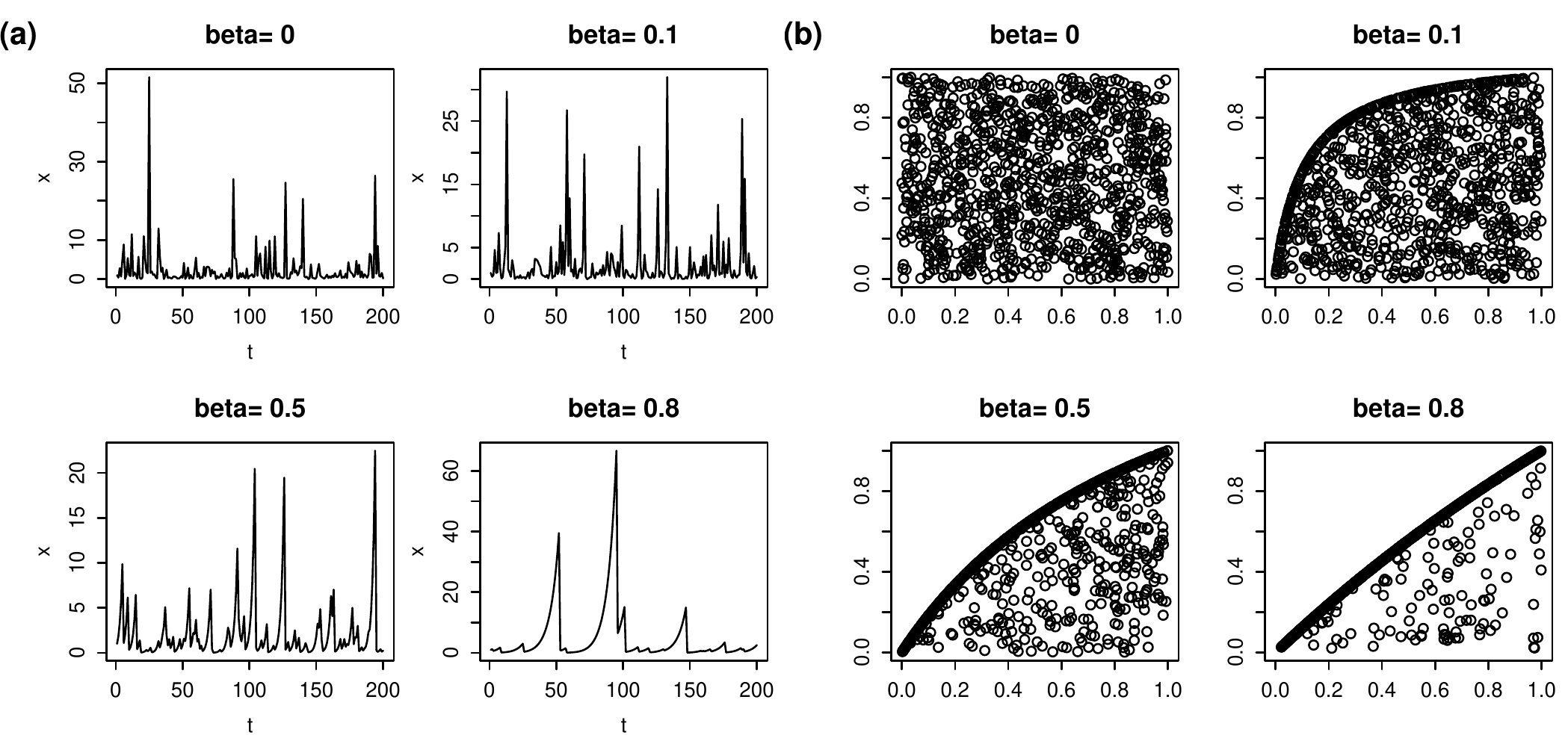}
}
\caption{\label{spangl:GPD} Shot-noise-type approach: (a) typical realizations,
and (b) lagged PP plots.}
\end{figure}

Figure\ \ref{spangl:GPD} shows typical realizations of the
shot-noise-type (SNT) approach, together with the corresponding lagged PP plots.
We refer to Desmettre\ et\ al.\ (2015) for the definition and theoretical foundations of this process as well as its application in liquidity risk management.

\subsection{Gaussian-copula approach}

Next, we propose to use a copula-based autoregressive model defined by
$$
X_t=G^{-1}\circ F\circ Y_t \ ,
$$
with
$$
Y_t = \rho Y_{t-1} + \bar \rho\, v_t \ ,
$$
where $\bar\rho^2 +\rho^2 = 1$, and
$v_t\stackrel{\text{iid}}{\sim} F={\cal N}(0,1)$.
Again, let $G$ and $G^{-1}$ denote the cdf and quantile
function of the ${\rm GPD}(\mu,\sigma,\xi)$.
Hence, $X_t\sim G$, and the parameter $\rho \in [-1, 1]$ controls the
dependency.
In the same way we may transform a more complex model, e.g., a Gaussian
ARMA-model to a GPD-process.
This again separates the estimation of time-dependency (Gaussian ARMA) and
of GPD-parameters, $\mu$, $\sigma$, $\xi$, of the marginal distribution.

\begin{figure}[bt!]\centering
\scalebox{0.35}{
\includegraphics{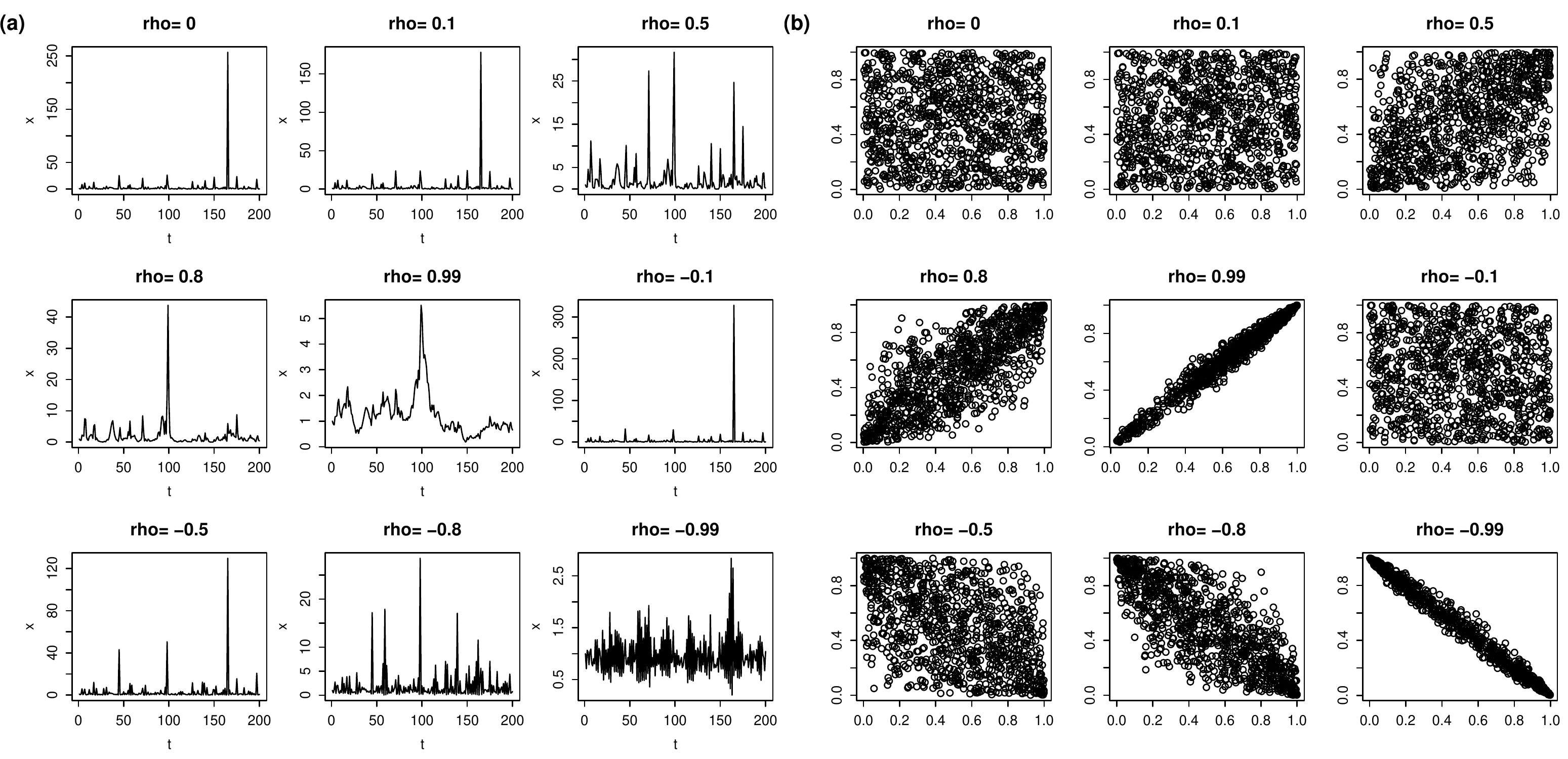}
}
\caption{\label{spangl:GPDGauss} Gaussian-copula approach: (a) typical realizations,
and (b) lagged PP plots.}
\end{figure}

In Figure\ \ref{spangl:GPDGauss} we see typical realizations of the
Gaussian-copula (GC) approach,
together with the corresponding lagged PP plots.

\subsection{GLM approach}

Last, we suggest a parameter driven approach. We define our GPD-process
by
$$
X_t \sim {\rm GPD}(\mu,\sigma_t,\xi_t) \ .
$$
The scale parameter $\sigma_t$ is given by
$$
\sigma_t=\ell_1(\beta_1+\beta_2 h_1(X_{t-1})) \ ,
$$
with $\ell_1$ and $h_1$ properly chosen, e.g.,
$\ell_1\colon \mathbb{R}\to\mathbb{R}_+$, and $h_1(x)=x^2$.

The shape parameter $\xi_t$ is given by
$$
\xi_t=\ell_2(\gamma_1+\gamma_2 h_2(X_{t-1})) \ ,
$$
with $\ell_2$ and $h_2$ again properly chosen, e.g.,
$\ell_2\colon \mathbb{R}\to(-0.5,2.5)$, and $h_2(x)=\log(1+|x|)$.
$h_2$ controls the tails.
Moreover, we note that it is essentially to chose the codomain of
$\ell_2$ equal to (-0.5,2.5).

\begin{figure}[bt!]\centering
\scalebox{0.35}{
\includegraphics{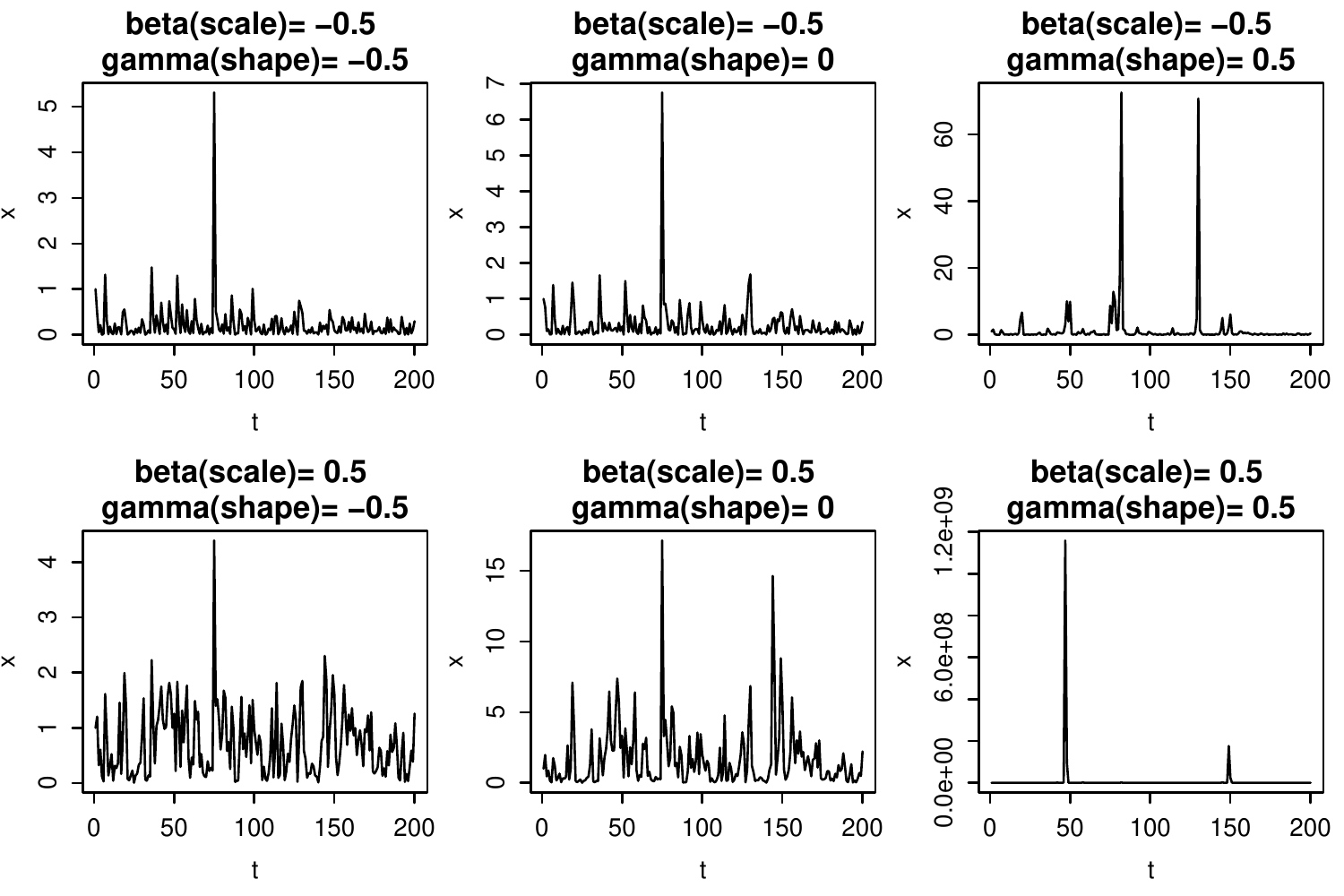}
}
\caption{\label{spangl:GPDGLM} GLM approach: typical realizations.}
\end{figure}

Typical realizations of the GLM approach are plotted in
Figure\ \ref{spangl:GPDGLM}.
Here, lagged PP plots do not make sense as the dependence
structure is modeled by fitting the GPD-parameters by GLMs
where previous observations enter as regressors.
See Pupashenko\ et\ al.\ (2014) for theoretical foundations.

\section{Evaluation and real world data set} 


To evaluate our four approaches we use daily average discharge data
of the Danube river at Donauw\"orth from 1978 to 2008.
%
%
We obtain corresponding
model fits and simulated data as visible in the following pictures. 

\begin{figure}[bt!]\centering
\scalebox{0.35}{
\includegraphics{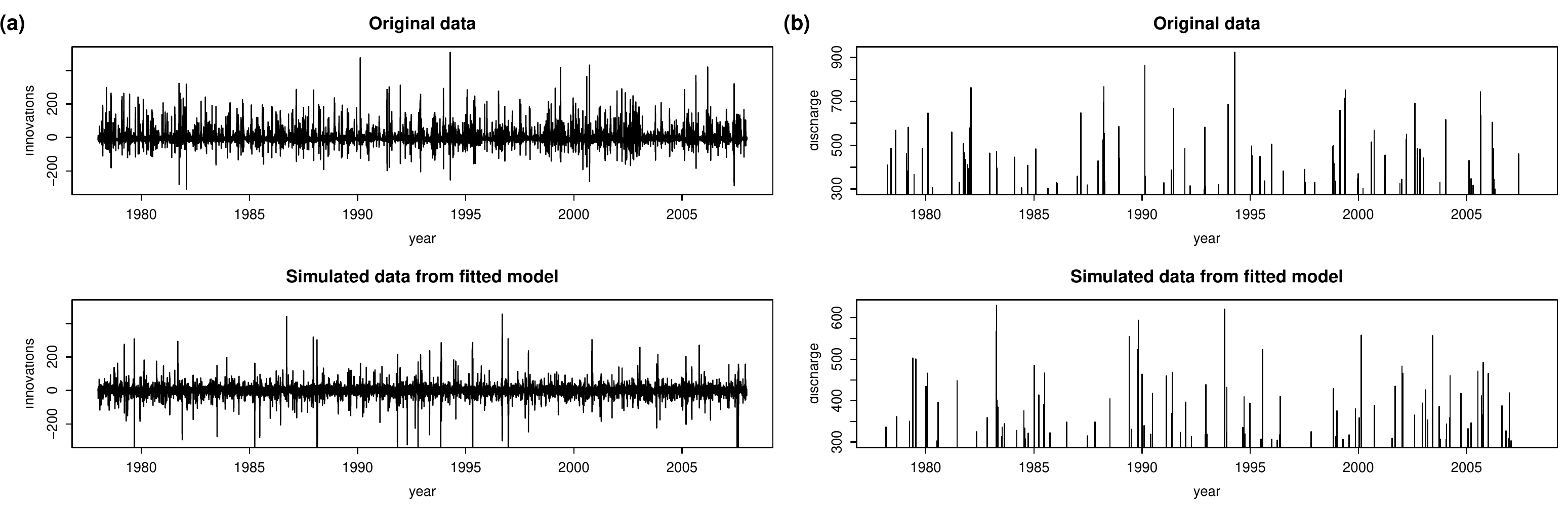}
}
\caption{\label{spangl:GPDSSMarGPD-data} Real data vs. simulation:
(a) SSM approach, and (b) shot-noise-type approach.}
\end{figure}

\begin{figure}[bt!]\centering
\scalebox{0.35}{
\includegraphics{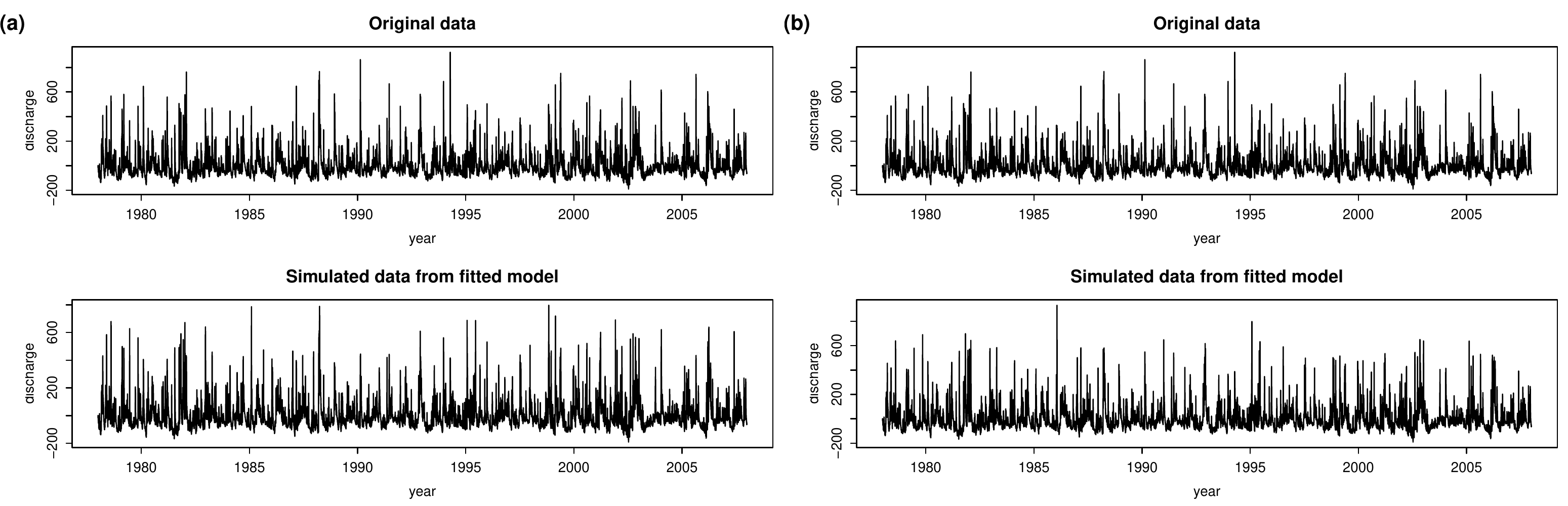}
}
\caption{\label{spangl:GPDGaussGLM-data} Real data vs. simulation:
(a) Gaussian-copula approach, and (b) GLM approach.}
\end{figure}

In Figures\ \ref{spangl:GPDSSMarGPD-data} and \ref{spangl:GPDGaussGLM-data}
the upper panels always show the original data and the lower panels one
simulated path.

Figure\ \ref{spangl:GPDSSMarGPD-data}a shows the innovations of the
original series after filtering and the simulated innovations using the
SSM approach. Here, instead of using an EARCH model as proposed in
Section 2.1, we use a GARCH(1,1) model with $t$-distributed innovations.
Comparing the two series in Figure\ \ref{spangl:GPDSSMarGPD-data}a
we see that in the lower panel the volatility clusters are reproduced
quite well. However, negative innovations are over-represented in
the simulated series.

In Figure\ \ref{spangl:GPDSSMarGPD-data}b only the extreme
events above a threshold of 300 of the original data as well as the simulated
ones using the shot-noise-type approach are plotted.
Comparing the two plots in Figure\ \ref{spangl:GPDSSMarGPD-data}b
we note that the distribution of extreme events, i.e., the inter-arrival
time distribution of extremes, coincide well, whereas the level of discharge is slightly under-estimated by the simulated process.

In the upper panels of Figure\ \ref{spangl:GPDGaussGLM-data} the original
detrended and deseasonalized series is displayed. The lower panel of
Figure\ \ref{spangl:GPDGaussGLM-data}a shows one simulated path using
the Gaussian-copula approach whereas in Figure\ \ref{spangl:GPDGaussGLM-data}b
we see a simulated path using the GLM approach. Comparing
the lower panels of
Figures\ \ref{spangl:GPDGaussGLM-data}a and \ref{spangl:GPDGaussGLM-data}b
with the upper ones we see that the level of discharge is estimated very
well by the Gaussian-copula and the GLM approach.
However, we note that using these two
approaches we are not able to assess the inter-arrival times.

\section{Summary and conclusion} 

We presented four flexible and parametric approaches to model dynamics
as well as extremes.
We remark that all models proposed are parsimonious ones.
Moreover, we are able to get grip on the inter-arrival
time distribution of extremes.
The SSM and GLM approaches are able to
model complex dynamics. The SNT as well as the SSM approach
are able to assess the inter-arrival time distribution of the extreme
events well. The GC approach can easily be extended to
capture multivariate dependency.


\acknowledgments{All authors gratefully acknowledge financial support
by the Volkswagen Foundation for the project ``Robust Risk Estimation'',
{\tt http://www.mathematik.uni-kl.de/}$\scriptstyle \sim${\tt wwwfm/RobustRiskEstimation}.}



\references
\begin{description}
%
%
\item[Desmettre, S., de Kock, J., Ruckdeschel, P. and Seifried, F.T.] (2015).
Generalized Pareto Processes and Liquidity.
{\it Working Paper\/}.
\item[Khaliq, M.N., Ouarda, T.B.M.J., Ondo, J.-C., Gachon, P.,
      and Bobee, B.] (2006).
Frequency analysis of a sequence of dependent and/or non-stationary
hydro-meteorological observations: A review.
{\it Journal of Hydrology\/}, {\bf 329}, 534\,--\,552.
\item[Pupashenko, D., Ruckdeschel, P., and Kohl, M.] (2015).
$L_2$ Differentiability of Generalized Linear Models.
\textit{Statist. Probab. Lett.} \textbf{97}: 155--164.
\end{description}

\end{document}